\documentclass[twocolumn,floatfix,showpacs,pre,superscriptaddress]{revtex4}%
\usepackage{graphicx,graphics,color,epsfig}
\usepackage{bm}
\usepackage{amsmath}
\usepackage{amssymb}
\usepackage{subfigure}
\usepackage{amsfonts}
\usepackage{graphicx}%
\providecommand{\U}[1]{\protect\rule{.1in}{.1in}}
\begin{document}
\title{Rigorous proof for the non-local correlation function \\in the antiferromagnetic seamed transverse Ising ring}
\author{Jian-Jun Dong}
\affiliation{College of Physical Science and Technology, Sichuan University, 610064,
Chengdu, P. R. China}
\affiliation{Key Laboratory of High Energy Density Physics and Technology of Ministry of
Education, Sichuan University, 610064, Chengdu, P. R. China}
\author{Zhen-Yu Zheng}
\affiliation{College of Physical Science and Technology, Sichuan University, 610064,
Chengdu, P. R. China}
\affiliation{Key Laboratory of High Energy Density Physics and Technology of Ministry of
Education, Sichuan University, 610064, Chengdu, P. R. China}
\author{Peng Li}
\email{lipeng@scu.edu.cn}
\affiliation{College of Physical Science and Technology, Sichuan University, 610064,
Chengdu, P. R. China}
\affiliation{Key Laboratory of High Energy Density Physics and Technology of Ministry of
Education, Sichuan University, 610064, Chengdu, P. R. China}
\date{\today}

\begin{abstract}
An unusual correlation function is conjectured by M. Campostrini \emph{et al}.
(Phys. Rev. E \textbf{{91}}, 042123 (2015)) for the ground state of a
transverse Ising chain with geometrical frustration in one of the translationally invariant cases. 
Later, we demonstrated the correlation function and showed its non-local
nature in the thermodynamic limit based on the rigorous evaluation of a
Toeplitz determinant (J. Stat. Mech. 113102 (2016)). In this paper, we prove
rigorously that all the states that forming the lowest gapless spectrum
(including the ground state) in the kink phase exhibit the same asymptotic
correlation function. So, in a point of view of cannonical ensemble,
the thermal correlation function is inert to temperature within the energy range of the lowest gapless spectrum.

\end{abstract}

\pacs{05.50.+q, 75.50.Ee, 02.30.Tb}
\maketitle



\section{Introduction}

Quantum antiferromagnetic spin chain is one of the subjects in quantum
magnetism endowed with rich and interesting physics \cite{McCoy1,Nagaosa}. In
recent years, spin chains with deliberate boundary conditions that induce geometrical frustration
has been getting more and more attention \cite{Solomon,Owerre,Yamanaka,Marzolono}. The simplest and
fundamental model providing many essential physics of interest may be the
transverse field Ising chain \cite{Lieb,Pfeuty,Suzuki,Sachdev,Dutta}
\begin{equation}
H_{s}=\sum_{j=1}^{N}J_{j}\sigma_{j}^{x}\sigma_{j+1}^{x}-\sum_{j=1}^{N}%
h_{j}\sigma_{j}^{z}, \label{HIsing}%
\end{equation}
with Pauli matrices $\sigma_{j}^{\alpha}$($\alpha=x,z$), exchange coupling
$J_{j}$ and external field $h_{j}$. When special boundary conditions are
imposed, unusual properties may emerge due to geometrical frustration of spin arrangement
\cite{Cabrera,Vicari1,Vicari2}. In treating the ferromagnetic transverse Ising chain seamed by
one antiferromagnetic bond, Campostrini \textit{et al}. \cite{Vicari1} conjectured an
usual correlation function for the ground state based on numerical
calculations. Later, in the context of translational invariance and the "a-cycle
problem" \cite{Lieb}, we analysed the antiferromagnetic seamed system with
$J_{j}=J>0$, $h_{j}=h$ and $N\in$ Odd \cite{Dong1}, which is equivalent to one
of the translational invariant case of the model considered by
Campostrini \textit{et al}. \cite{Vicari1}. And based on rigorous solutions, we proved the
conjectured correlation function, which exhibits a non-local nature in the
thermodynamic limit \cite{Dong1}.

Nonetheless, as disclosed by the rigorous solutions,
there are $2N-1$ similar low-lying quantum states above the ground state
\cite{Dong1}, which form a gapless spectrum in the kink phase ($h<J$) in the thermodynamic limit \cite{Vicari1}. In this paper, we prove rigorously that the correlation
functions of all the $2N$ quantum states (including the ground state) in the
gapless spectrum have the same unusual asymptotic behaviour in the thermodynamic limit.

The organization of the paper is as follows. In Sec. II, the low-lying $2N$
quantum states of the a-cycle problem with ring frustration are reviewed. In
Sec. III, we show how the correlation functions are linked to a special type of Toeplitz
determinant. In Sec. IV, we give the rigorous proof of a generalized theorem for evaluating of the Toeplitz determinant. At last, the unusual correlation function is worked out in Sec. V.

\section{The low-lying $2N$ quantum states}

Here we briefly review the low-lying $2N$ quantum states of the frustrated
ring that we obtained in our previous work \cite{Dong1}.

First, the spin ring is mapped to a model of spinless fermions by the
Jordan-Wigner transformation \cite{J-W}
\begin{equation}
\sigma_{j}^{+}=\left(  \sigma_{j} ^{x}+\operatorname{i}\sigma_{j}^{y}\right)
/2=c_{j}^{\dag}\exp( \operatorname{i}\pi\sum_{l<j}c_{l}^{\dag}c_{l}).
\end{equation}
Then Eq. (\ref{HIsing}) takes the form
\begin{align}
H_{f}  &  =Nh-2h\sum_{j=1}^{N}c_{j}^{\dag}c_{j}+J\sum_{j=1}^{N-1}(c_{j}^{\dag
}-c_{j})(c_{j+1}^{\dag}-c_{j+1})\nonumber\\
&  \quad-J\exp(\operatorname{i}\pi M)(c_{N}^{\dag}-c_{N})(c_{1}^{\dag}+c_{1}),
\label{H2}%
\end{align}
where $M=\sum_{l=1}^{N}c_{l}^{\dag}c_{l}$. It has been shown that the last
term in Eq. (\ref{H2}) can not be discarded for the frustrated ring, i.e.
$J>0$ and $N\in$ odd. And the system should be solved faithfully in the
framework of "a-cycle problem" \cite{Dong1,Dong2}. In the fermion
representation, the Hilbert space of the Hamiltonian, Eq. (\ref{H2}), is
enlarged twice as the one of the original spin Hamiltonian, Eq. (\ref{HIsing}%
). The quantum states of Eq. (\ref{H2}) are grouped into two channels, the odd
and even channels labelled by $M\in$ odd and $M\in$ even respectively. To
restore the full degrees of freedom (DOF) of the original spin Hamiltonian, we
should project out half redundant DOF in each channel \cite{Lieb2}.

Second, the Hamiltonian Eq. (\ref{H2}) is easily solved with the help of
parity constraint, Fourier transformation $c_{q}$=$\frac{1}{\sqrt{N}}%
\sum_{j=1}^{N}c_{j}\exp\left(  \operatorname{i}q\,j\right)  $ and Bogoliubov
transformation ($q\neq0$ and $\pi$)%
\begin{equation}
\eta_{q}=u_{q}c_{q}-\operatorname{i}v_{q}c_{-q}^{\dagger}%
\end{equation}
with
\begin{align}
u_{q}^{2}  &  =\frac{1}{2}\left(  1+\frac{\epsilon(q)}{\omega(q)}\right)
,v_{q}^{2}=\frac{1}{2}\left(  1-\frac{\epsilon(q)}{\omega(q)}\right)
,\nonumber\\
2u_{q}v_{q}  &  =\frac{\Delta(q)}{\omega(q)},\epsilon(q)=J\cos{q}%
-h,\Delta(q)=J\sin{q,}\nonumber\\
\omega(q)  &  =\sqrt{h^{2}+J^{2}-2hJ\cos{q}}.
\end{align}
The momentum values $q=0$ and $\pi$ play an important role in controlling the
parity of valid spin states. In momentum space, the spin Hamiltonian can be
expressed as \cite{Dziarmaga,Fabio}
\begin{equation}
H_{s}=PH_{f}^{(e)}P\oplus PH_{f}^{(o)}P,
\end{equation}
where $PH_{f}^{(e)}P$ means only even number fermionic occupation states are
valid states for the original spin model due to the even parity constraint,
while $PH_{f}^{(o)}H$ implies that only odd number fermionic occupation states
are valid states, and $P$ means projection of the redundant DOF \cite{Lieb2}.
$H_{f}^{(e)}$ and $H_{f}^{(o)}$ are given by
\begin{equation}
H_{f}^{(e)}=\epsilon(\pi)\left(  2c_{\pi}^{\dagger}c_{\pi}-1\right)
+\sum_{q\in q^{(e)},q\neq\pi}\omega(q)\left(  2\eta_{q}^{\dagger}\eta
_{q}-1\right)  , \label{He}%
\end{equation}%
\begin{equation}
H_{f}^{(o)}=\epsilon(0)\left(  2c_{0}^{\dagger}c_{0}-1\right)  +\sum_{q\in
q^{(o)},q\neq0}\omega(q)\left(  2\eta_{q}^{\dagger}\eta_{q}-1\right)  ,
\label{Ho}%
\end{equation}
with
\begin{equation}
q^{(e)}=\{-\frac{N-2}{N}\pi,\ldots,-\frac{1}{N}\pi,\frac{1}{N}\pi,\ldots
,\frac{N-2}{N}\pi,\pi\},
\end{equation}%
\begin{equation}
q^{(o)}=\{-\frac{N-1}{N}\pi,\ldots,-\frac{2}{N}\pi,0,\frac{2}{N}\pi
,\ldots,\frac{N-1}{N}\pi\}.
\end{equation}

In the kink phase ($h<J$), it is easy to write down the $2N$ low-lying states
of the gapless spectrum of width $4h$. The ground state comes from the odd
channel and reads%
\begin{equation}
|E_{0}^{(o)}\rangle=c_{0}^{\dag}|\phi^{(o)}\rangle. \label{E0}%
\end{equation}
The upper-most state comes from the even channel and reads%
\begin{equation}
|E_{\pi}^{(e)}\rangle=|\phi^{(e)}\rangle. \label{Epi}%
\end{equation}
The rest $2(N-1)$ states interweave between $|E_{0}\rangle$ and $|E_{\pi
}\rangle$. Half of them are of odd parity
\begin{equation}
|E_{k}^{(o)}\rangle=\eta_{k}^{\dag}|\phi^{(o)}\rangle,\left\{  k\in
q^{(o)}|k\neq0\right\}  ,
\end{equation}
and half of them are of even parity%
\begin{equation}
|E_{k}^{(e)}\rangle=\eta_{k}^{\dag}c_{\pi}^{\dag}|\phi^{(e)}\rangle,\left\{
k\in q^{(e)}|k\neq\pi\right\}  .
\end{equation}
In the above, the BCS-type states
\begin{align}
|\phi^{(o)}\rangle &  =\prod_{\substack{q\in q^{(o)},0<q<\pi}}\left(
u_{q}+\operatorname{i}v_{q}c_{q}^{\dag}c_{-q}^{\dag}\right)  |0\rangle,\\
|\phi^{(e)}\rangle &  =\prod_{\substack{q\in q^{(e)},0<q<\pi}}\left(
u_{q}+\operatorname{i}v_{q}c_{q}^{\dag}c_{-q}^{\dag}\right)  |0\rangle,
\end{align}
are vacuums corresponding to $H_{f}^{(e)}$ and $H_{f}^{(o)}$ respectively, and
$\left\vert 0\right\rangle =\left\vert \downarrow\downarrow\downarrow
\cdots\downarrow\right\rangle $.

There is a rough but nice picture for the low-lying states in a pertubative
treatment, which gives the following superposed states \cite{Dong1},%
\begin{align}
|A_{p}\rangle &  =\frac{1}{\sqrt{2N}}\sum_{j}e^{-\operatorname{i}%
p\,j}(|K(j),\leftarrow\rangle+|K(j),\rightarrow\rangle),\label{Ak}\\
|B_{p}\rangle &  =\frac{1}{\sqrt{2N}}\sum_{j}e^{-\operatorname{i}%
p\,j}(|K(j),\leftarrow\rangle-|K(j),\rightarrow\rangle),\label{Bk}\\
p  &  =\{-\frac{N-1}{N}\pi,\ldots,-\frac{2}{N}\pi,0,\frac{2}{N}\pi
,\ldots,\frac{N-1}{N}\pi\}.
\end{align}
They are translational invariant states composed of the classical kink states,%
\begin{align}
|K(j),  &  \rightarrow\rangle=|...,\leftarrow_{j-1},\rightarrow_{j}%
,\rightarrow_{j+1},\leftarrow_{j+2},...\rangle,\nonumber\\
|K(j),  &  \leftarrow\rangle=|...,\rightarrow_{j-1},\leftarrow_{j}%
,\leftarrow_{j+1},\rightarrow_{j+2},...\rangle.
\end{align}
One can find the correspondence between the exact states and the approximate
states ($h\ll J$)
\begin{align}
|E_{0}^{(o)}\rangle &  \approx|A_{0} \rangle, \quad|E_{\pi}^{(e)}%
\rangle\approx|B_{0}\rangle,\\
|E_{q}^{(o)}\rangle &  \approx|A_{q} \rangle,\quad|E_{q}^{(e)}\rangle
\approx|B_{\pi-q}\rangle.
\end{align}

\section{Longitudinal correlation functions}

The two point longitudinal spin-spin correlation function for arbitrary state
$|\psi\rangle$ is defined as
\begin{equation}
C_{r,N}^{xx}(|\psi\rangle)=\langle\psi|\sigma_{j}^{x}\sigma_{j+r}^{x}%
|\psi\rangle.
\end{equation}
Due to the translational invariance, the correlation function depends on the
separation $r$ between two spins rather than the lattice position $j$. And
because of the periodic boundary condition, we have a cyclic relation,
$C_{r,N}^{xx}(|\psi\rangle)=C_{N-r,N}^{xx}(|\psi\rangle)$, for a
translationally invariant state $|\psi\rangle$.

By the approximate states, Eq. (\ref{Ak})-(\ref{Bk}), we can easily get%
\begin{equation}
C_{r,N}^{xx}(|A_{p}\rangle)=C_{r,N}^{xx}(|B_{p}\rangle)=(-1)^{r}(1-\frac{2r}{N}).
\label{Cxxper}%
\end{equation}
When approaching the thermodynamic limit $N\rightarrow
\infty$, one can set $r/N\rightarrow0$ to see a local correlation or set a
finite $r/N$ to see a non-local correlation. Since the approximate states
are valid for $h\ll J$, so does Eq. (\ref{Cxxper}).

A rigorous result for the exact ground state has been proved in our previous
work, which reads ($N\gg1$)%
\begin{equation}
C_{r,N}^{xx}(|E_{0}^{(o)}\rangle)=(-1)^{r}(1-\frac{h^{2}}{J^{2}}%
)^{1/4}(1-\frac{2r}{N}). \label{CxxE0}%
\end{equation}
The same result was conjectured by Campostrini \textit{et al}. in a context of
the transverse Ising ring with one-bond defect \cite{Vicari1}. Now we
demonstrate that the rest $2N-1$ low-lying states exhibit the same behavior.


First, we show that all of the correlation functions of the $2N$ low-lying
states can be casted into a general Toeplitz determinant. Comparing with the
case of ground state, the general Toeplitz determinant is much more
complicated and depends on the wave number $k$. We should also be careful to
the parity channels.

\begin{widetext}
The correlation function of the ground state $|E_{0}^{(o)}\rangle=c_{0}^{\dag
}|\phi^{(o)}\rangle$ is given by a determinant as
\begin{align}
C_{r,N}^{xx}\left(  |E_{0}^{(o)}\rangle\right)   &  =\langle\phi^{(o)}%
|c_{0}B_{j}A_{j+1}\ldots B_{j+r-1}A_{j+r}c_{0}^{\dag}|\phi^{(o)}%
\rangle\nonumber\\
&  =\left\vert
\begin{array}
[c]{cccc}%
D_{0}+\frac{2}{N} & D_{-1}+\frac{2}{N} & \cdots & D_{1-r}+\frac{2}{N}\\
D_{1}+\frac{2}{N} & D_{0}+\frac{2}{N} & \cdots & D_{2-r}+\frac{2}{N}\\
\cdots & \cdots & \cdots & \cdots\\
D_{r-1}+\frac{2}{N} & D_{r-2}+\frac{2}{N} & \cdots & D_{0}+\frac{2}{N}%
\end{array}
\right\vert ,\label{corgs}%
\end{align}
where, following Lieb \cite{Lieb}, we have introduced $A_{j}=c_{j}^{\dag
}+c_{j}$ and $B_{j}=c_{j}^{\dag}-c_{j}$. And we have making use of the Wick's
theorem and the contractions in respect of $|\phi^{(o)}\rangle$: $\langle
A_{l}A_{m}\rangle=\delta_{lm}$, $\langle c_{0}c_{0}^{\dag} \rangle=1$, $\langle B_{l}B_{m}\rangle=-\delta_{lm}$,
$\langle A_{j}c_{0}^{\dag}\rangle=-\langle B_{j}c_{0}^{\dag}\rangle=\frac
{1}{\sqrt{N}}$, $\langle B_{l}A_{m}\rangle=D_{l-m+1}$,  with $D_{n}$=$\frac
{1}{N}\sum_{_{\substack{q\in q^{(o)}}}}D(\operatorname{e}^{\operatorname{i}%
q})\exp\left(  -\operatorname*{i}q\,n\right)  $ and%
\begin{equation}
D(\operatorname{e}^{\operatorname{i}q})=-\frac{J-h\operatorname{e}%
^{-\operatorname{i}q}}{\sqrt{\left(  J-h\operatorname{e}^{-\operatorname{i}%
q}\right)  \left(  J-h\operatorname{e}^{\operatorname{i}q}\right)  }}.
\end{equation}
For the $N-1$ states from odd parity $|E_{k}^{(o)}\rangle=\eta_{k}^{\dag}%
|\phi^{(o)}\rangle$, $\left\{  k\in q^{(o)}|k\neq0\right\}  $, we can arrive
at
\begin{align}
C_{r,N}^{xx}\left(  |E_{k}^{(o)}\rangle\right)   &  =\langle\phi^{(o)}%
|\eta_{k}B_{j}A_{j+1}\ldots B_{j+r-1}A_{j+r}\eta_{k}^{\dagger}|\phi
^{(o)}\rangle\nonumber\\
&  =\frac{1}{2}\left[  \Gamma^{(o)}\left(  r,N,\alpha_{k},\operatorname{e}%
^{\operatorname{i}k}\right)  +\Gamma^{(o)}\left(  r,N,\alpha_{-k}%
,\operatorname{e}^{-\operatorname{i}k}\right)  \right]  ,\label{corodd}%
\end{align}
where%
\begin{equation}
\Gamma^{(o)}(r,N,\alpha_{k},\operatorname{e}^{\operatorname*{i}k})=\left\vert
\begin{array}
[c]{cccc}%
D_{0}+\frac{2\alpha_{k}}{N} & D_{-1}+\frac{2\alpha_{k}}{N}\operatorname{e}%
^{-\operatorname*{i}k} & \cdots & D_{1-r}+\frac{2\alpha_{k}}{N}%
\operatorname{e}^{\operatorname*{i}(1-r)k}\\
D_{1}+\frac{2\alpha_{k}}{N}\operatorname{e}^{\operatorname*{i}k} & D_{0}%
+\frac{2\alpha_{k}}{N} & \cdots & D_{2-r}+\frac{2\alpha_{k}}{N}%
\operatorname{e}^{\operatorname*{i}(2-r)k}\\
\cdots & \cdots & \cdots & \cdots\\
D_{r-1}+\frac{2\alpha_{k}}{N}\operatorname{e}^{\operatorname*{i}(r-1)k} &
D_{r-2}+\frac{2\alpha_{k}}{N}\operatorname{e}^{\operatorname*{i}(r-2)k} &
\cdots & D_{0}+\frac{2\alpha_{k}}{N}%
\end{array}
\right\vert ,\label{Gammao}%
\end{equation}
and $\langle\eta_{k}B_{l}\rangle\langle A_{m}\eta_{k}^{\dagger}\rangle$%
=$\frac{\alpha_{k}}{N}\exp\left(  \operatorname*{i}k\,(l-m+1)\right)  ,$
$\alpha_{k}\equiv-D(\operatorname{e}^{\operatorname{i}k}).$

Next, we consider the $N-1$ states from even parity $|E_{k}^{(e)}\rangle
=\eta_{k}^{\dag}c_{\pi}^{\dagger}|\phi^{(e)}\rangle$, $\left\{  k\in
q^{(e)}|k\neq\pi\right\}  $. Similarly, Wick's theorem can be applied and we
can arrive at
\begin{align}
C_{r,N}^{xx}\left(  |E_{k}^{(e)}\rangle\right)   &  =\langle\phi^{(e)}|c_{\pi
}\eta_{k}B_{j}A_{j+1}\ldots B_{j+r-1}A_{j+r}\eta_{k}^{\dagger}c_{\pi}%
^{\dagger}|\phi^{(e)}\rangle\nonumber\\
&  =\frac{1}{2}\left[  \Gamma^{(e)}(r,N,\alpha_{k},\operatorname{e}%
^{\operatorname*{i}k})+\Gamma^{(e)}(r,N,\alpha_{-k},\operatorname{e}%
^{-\operatorname*{i}k})\right]  ,\label{coreven}%
\end{align}
where%
\begin{equation}
\Gamma^{(e)}(r,N,\alpha_{k},\operatorname{e}^{\operatorname*{i}k})=\left\vert
\begin{array}
[c]{cccc}%
F_{0}+\frac{2\alpha_{k}}{N} & F_{-1}+\frac{2\alpha_{k}}{N}\operatorname{e}%
^{-\operatorname*{i}k} & \cdots & F_{1-r}+\frac{2\alpha_{k}}{N}%
\operatorname{e}^{\operatorname*{i}(1-r)k}\\
F_{1}+\frac{2\alpha_{k}}{N}\operatorname{e}^{\operatorname*{i}k} & F_{0}%
+\frac{2\alpha_{k}}{N} & \cdots & F_{2-r}+\frac{2\alpha_{k}}{N}%
\operatorname{e}^{\operatorname*{i}(2-r)k}\\
\cdots & \cdots & \cdots & \cdots\\
F_{r-1}+\frac{2\alpha_{k}}{N}\operatorname{e}^{\operatorname*{i}(r-1)k} &
F_{r-2}+\frac{2\alpha_{k}}{N}\operatorname{e}^{\operatorname*{i}(r-2)k} &
\cdots & F_{0}+\frac{2\alpha_{k}}{N}%
\end{array}
\right\vert ,
\end{equation}%
\begin{equation}
F_{n}=\frac{1}{N}\sum_{_{\substack{q\in q^{(e)}}}}\exp\left(
-\operatorname*{i}q\,n\right)  \frac{-\left(  J-h\operatorname{e}%
^{\operatorname{i}q}\right)  }{\sqrt{\left(  J-h\operatorname{e}%
^{-\operatorname{i}q}\right)  \left(  J-h\operatorname{e}^{\operatorname{i}%
q}\right)  }}.
\end{equation}
Finally, for the upper-most level $|E_{\pi}^{(e)}\rangle=|\phi^{(e)}\rangle$,
after the contractions with $|\phi^{(e)}\rangle$ we arrive at
\begin{align}
C_{r,N}^{xx}\left(  |E_{\pi}^{(e)}\rangle\right)   &  =\langle\phi^{(e)}%
|B_{j}A_{j+1}\ldots A_{j+r-1}B_{j+r-1}A_{j+r}|\phi^{(e)}\rangle\nonumber\\
&  =\left\vert
\begin{array}
[c]{cccc}%
F_{0}+\frac{2}{N} & F_{-1}+\frac{2}{N}\operatorname{e}^{-\operatorname*{i}\pi}
& \cdots & F_{1-r}+\frac{2}{N}\operatorname{e}^{\operatorname*{i}\pi\left(
1-r\right)  }\\
F_{1}+\frac{2}{N}\operatorname{e}^{\operatorname*{i}\pi} & F_{0}+\frac{2}{N} &
\cdots & F_{2-r}+\frac{2}{N}\operatorname{e}^{\operatorname*{i}\pi\left(
2-r\right)  }\\
\cdots & \cdots & \cdots & \cdots\\
F_{r-1}+\frac{2}{N}\operatorname{e}^{\operatorname*{i}\pi\left(  r-1\right)  }
& F_{r-2}+\frac{2}{N}\operatorname{e}^{\operatorname*{i}\pi\left(  r-2\right)
} & \cdots & F_{0}+\frac{2}{N}%
\end{array}
\right\vert .\label{corpi}%
\end{align}
For large enough $N$, we have%
\begin{equation}
D_{n}=F_{n}=\int_{-\pi}^{\pi}\frac{dq}{2\pi}\operatorname{e}%
^{-\operatorname*{i}qn}\frac{-\left(  J-h\operatorname{e}^{\operatorname*{i}%
q}\right)  }{\sqrt{\left(  J-h\operatorname{e}^{-\operatorname*{i}q}\right)
\left(  J-h\operatorname{e}^{\operatorname*{i}q}\right)  }}.
\end{equation}
So the correlation functions for the $2N$ low-lying states, Eq. (\ref{corgs}),
Eq. (\ref{corodd}), Eq. (\ref{coreven}) and Eq. (\ref{corpi}), can be written
in a uniform formula in the thermodynamic limit,%
\begin{equation}
C_{r,N}^{xx}\left(  |E_{k}^{(o/e)}\rangle\right)  =\frac{1}{2}\left[
\Gamma(r,N,\alpha_{k},\operatorname{e}^{\operatorname*{i}k})+\Gamma
(r,N,\alpha_{-k},\operatorname{e}^{-\operatorname*{i}k})\right]  ,
\end{equation}
where
\begin{equation}
\Gamma(r,N,\alpha_{k},\operatorname{e}^{\operatorname*{i}k})=\left\vert
\begin{array}
[c]{cccc}%
\overset{\sim}{D}_{0} & \overset{\sim}{D}_{-1} & \cdots & \overset{\sim
}{D}_{1-r}\\
\overset{\sim}{D}_{1} & \overset{\sim}{D}_{0} & \cdots & \overset{\sim
}{D}_{2-r}\\
\cdots & \cdots & \cdots & \cdots\\
\overset{\sim}{D}_{r-1} & \overset{\sim}{D}_{r-2} & \cdots & \overset{\sim
}{D}_{0}%
\end{array}
\right\vert ,\label{Gamma}%
\end{equation}%
\begin{equation}
\overset{\sim}{D}_{n}=\int_{-\pi}^{\pi}\frac{dq}{2\pi}\,\operatorname{e}%
^{-\operatorname*{i}qn}\frac{-\left(  J-h\operatorname{e}^{\operatorname*{i}%
q}\right)  }{\sqrt{\left(  J-h\operatorname{e}^{-\operatorname*{i}q}\right)
\left(  J-h\operatorname{e}^{\operatorname*{i}q}\right)  }}+\frac{2\alpha_{k}%
}{N}\operatorname{e}^{\operatorname*{i}k\,n},\label{Dn}%
\end{equation}
\end{widetext}

\section{Generalized theorem}

Due to the presence of the extra term $\frac{2\alpha_{k}}{N}\operatorname{e}%
^{\operatorname*{i}k\,n}$ in each element of the new Toeplitz determinant, Eq.
(\ref{Gamma}), the Szeg\"{o} limit theorem cannot applied directly
\cite{Silbermann,McCoy}. To evaluate the correlation functions analytically,
we need to prove a generalized theorem for this special determinant.

\textbf{Theorem: }\textit{Consider a Toeplitz determinant}%
\begin{equation}
\Theta(r,N,x,\operatorname{e}^{\operatorname*{i}k})=\left\vert
\begin{array}
[c]{cccc}%
\overset{\sim}{D}_{0} & \overset{\sim}{D}_{-1} & \cdots & \overset{\sim
}{D}_{1-r}\\
\overset{\sim}{D}_{1} & \overset{\sim}{D}_{0} & \cdots & \overset{\sim
}{D}_{2-r}\\
\cdots & \cdots & \cdots & \cdots\\
\overset{\sim}{D}_{r-1} & \overset{\sim}{D}_{r-2} & \cdots & \overset{\sim
}{D}_{0}%
\end{array}
\right\vert \label{sTheta}%
\end{equation}
\textit{with }%
\begin{equation}
\overset{\sim}{D}_{n}\mathit{=}\int_{-\pi}^{\pi}\frac{dq}{2\pi}%
\,D(\operatorname{e}^{\operatorname*{i}q})\,\operatorname{e}%
^{-\operatorname*{i}qn}+\frac{x}{N}\operatorname{e}^{\operatorname*{i}kn}.
\end{equation}
\textit{If the generating function }$D(\operatorname{e}^{\operatorname*{i}q}%
)$\textit{ and }$\ln D(\operatorname{e}^{\operatorname*{i}q})$\textit{\ are
continuous on the unit circle }$\left\vert \operatorname{e}^{\operatorname*{i}%
q}\right\vert =1$\textit{, then the behavior for large }$N$\textit{ of
}$\Theta(r,N,x,\operatorname{e}^{\operatorname*{i}k})$\textit{ is given by
(}$1\ll r<N$\textit{) }%
\begin{equation}
\Theta(r,N,x,\operatorname{e}^{\operatorname*{i}k})=\Delta_{r}\left(
1+\frac{xr}{ND\left(  \operatorname{e}^{-\operatorname*{i}k}\right)  }\right)
,
\end{equation}
\textit{where }%
\begin{align}
\Delta_{r}  &  =\mu^{r}\exp(\sum_{n=1}^{\infty}nd_{-n}d_{n}),\label{sdeltar}\\
\mu &  =\exp\left[  \int_{-\pi}^{\pi}\frac{dq}{2\pi}\,\ln D(\operatorname{e}%
^{\operatorname*{i}q})\right]  ,\\
d_{n}  &  =\int_{-\pi}^{\pi}\frac{dq}{2\pi}\,e^{-\operatorname*{i}qn}\ln
D(\operatorname{e}^{\operatorname*{i}q}),
\end{align}
\textit{if the sum in Eq. (\ref{sdeltar}) converges.}

\textbf{Proof: }Let\textit{\ }$\operatorname{e}^{\operatorname*{i}q}=\xi,$
$D_{n}=\int_{-\pi}^{\pi}\frac{dq}{2\pi}\,D(\xi)\xi^{-n}\,,$ then
$\overset{\sim}{D}_{n}=D_{n}+\frac{x}{N}\operatorname{e}^{\operatorname*{i}%
kn}.$ First, we rewrite Eq. (\ref{sTheta}) as\begin{widetext}
\begin{align*}
\Theta(r,N,x,\operatorname{e}^{\operatorname*{i}k})  &  =\left\vert
\begin{array}
[c]{cccc}%
D_{0} & D_{-1} & \cdots & D_{-r+1}\\
D_{1} & D_{0} & \cdots & D_{-r+2}\\
\cdots & \cdots & \cdots & \cdots\\
D_{r-1} & D_{r-2} & \cdots & D_{0}%
\end{array}
\right\vert +\left\vert
\begin{array}
[c]{cccc}%
\frac{x}{N} & D_{-1} & \cdots & D_{1-r}\\
\frac{x}{N}\operatorname{e}^{\operatorname*{i}k} & D_{0} & \cdots & D_{2-r}\\
\cdots & \cdots & \cdots & \cdots\\
\frac{x}{N}\operatorname{e}^{\operatorname*{i}(r-1)k} & D_{r-2} & \cdots &
D_{0}%
\end{array}
\right\vert \\
&  +\ldots+\left\vert
\begin{array}
[c]{cccc}%
D_{0} & \frac{x}{N}\operatorname{e}^{-\operatorname*{i}k} & \cdots & D_{2-r}\\
D_{1} & \frac{x}{N} & \cdots & D_{2-r}\\
\cdots & \cdots & \cdots & \cdots\\
D_{r-1} & \frac{x}{N}\operatorname{e}^{\operatorname*{i}(r-2)k} & \cdots &
D_{0}%
\end{array}
\right\vert +\left\vert
\begin{array}
[c]{cccc}%
D_{0} & D_{-1} & \cdots & \frac{x}{N}\operatorname{e}^{\operatorname*{i}%
(1-r)k}\\
D_{1} & D_{0} & \cdots & \frac{x}{N}\operatorname{e}^{\operatorname*{i}%
(2-r)k}\\
\cdots & \cdots & \cdots & \cdots\\
D_{r-1} & D_{r-2} & \cdots & \frac{x}{N}%
\end{array}
\right\vert .
\end{align*}
\end{widetext}Then we compose a set of linear equations
\begin{equation}
\sum_{m=0}^{r-1}D_{n-m}x_{m}^{(r-1)}=\frac{x}{N}\mathrm{e}^{\mathrm{i}%
kn}\;,\;0\leq n\leq r-1. \label{sumeq}%
\end{equation}
These equations have an unique solution for $x_{n}^{(r-1)}$ if there exists a
non-zero determinant:
\begin{equation}
\Delta_{r}\equiv\left\vert
\begin{array}
[c]{cccc}%
D_{0} & D_{-1} & \cdots & D_{1-r}\\
D_{1} & D_{0} & \cdots & D_{2-r}\\
\cdots & \cdots & \cdots & \cdots\\
D_{r-1} & D_{r-2} & \cdots & D_{0}%
\end{array}
\right\vert \neq0.
\end{equation}
By Cramer's rule, we have the solution:
\begin{align}
&  x_{0}^{(r-1)}=\frac{\left\vert
\begin{array}
[c]{cccc}%
\frac{x}{N} & D_{-1} & \cdots & D_{1-r}\\
\frac{x}{N}\operatorname{e}^{\operatorname*{i}k} & D_{0} & \cdots & D_{2-r}\\
\cdots & \cdots & \cdots & \cdots\\
\frac{x}{N}\operatorname{e}^{\operatorname*{i}(r-1)k} & D_{r-2} & \cdots &
D_{0}%
\end{array}
\right\vert }{\Delta_{r}}\;,\\
&  x_{1}^{(r-1)}=\frac{\left\vert
\begin{array}
[c]{cccc}%
D_{0} & \frac{x}{N}\operatorname{e}^{-\operatorname*{i}k} & \cdots & D_{2-r}\\
D_{1} & \frac{x}{N} & \cdots & D_{2-r}\\
\cdots & \cdots & \cdots & \cdots\\
D_{r-1} & \frac{x}{N}\operatorname{e}^{\operatorname*{i}(r-2)k} & \cdots &
D_{0}%
\end{array}
\right\vert }{\Delta_{r}}\;,\\
&  \qquad\qquad\qquad\qquad\qquad\vdots\nonumber\\
&  x_{r-1}^{(r-1)}=\frac{\left\vert
\begin{array}
[c]{cccc}%
D_{0} & D_{-1} & \cdots & \frac{x}{N}\operatorname{e}^{\operatorname*{i}%
(1-r)k}\\
D_{1} & D_{0} & \cdots & \frac{x}{N}\operatorname{e}^{\operatorname*{i}%
(2-r)k}\\
\cdots & \cdots & \cdots & \cdots\\
D_{r-1} & D_{r-2} & \cdots & \frac{x}{N}%
\end{array}
\right\vert }{\Delta_{r}}.
\end{align}
So we arrive at
\begin{equation}
\Theta(r,N,x,\operatorname{e}^{\operatorname*{i}k})=\Delta_{r}+\Delta_{r}%
\sum_{n=0}^{r-1}\operatorname{e}^{-\operatorname*{i}kn}x_{n}^{(r-1)}.
\label{sumxn}%
\end{equation}
For our problem, $\Delta_{r}$ can be evaluated directly by using Szeg\"{o}'s
theorem, so we need to know how to calculate the second term in Eq.
(\ref{sumxn}). Follow the standard Wiener-Hopf procedure \cite{McCoy Wu,McCoy,
Wu}, we consider a generalization of Eq. (\ref{sumeq})
\begin{equation}
\sum_{m=0}^{r-1}D_{n-m}x_{m}=y_{n}\text{, \ }0\leq n\leq r-1 \label{Wiener}%
\end{equation}
and define
\begin{equation}
x_{n}=y_{n}=0\quad\text{for}\quad n\leq-1\quad\text{and}\quad n\geq r
\end{equation}%
\begin{align}
v_{n}  &  =\sum_{m=0}^{r-1}D_{-n-m}x_{m}\quad\text{for}\quad n\geq1\nonumber\\
&  =0\quad\text{for}\quad n\leq0
\end{align}%
\begin{align}
u_{n}  &  =\sum_{m=0}^{r-1}D_{r-1+n-m}x_{m}\quad\text{for}\quad n\geq
1\nonumber\\
&  =0\quad\text{for}\quad n\leq0
\end{align}
We further define
\begin{align}
D\left(  \xi\right)   &  =\sum_{n=-\infty}^{\infty}D_{n}\xi^{n},\quad Y\left(
\xi\right)  =\sum_{n=0}^{r-1}y_{n}\xi^{n},\nonumber\\
V\left(  \xi\right)   &  =\sum_{n=1}^{\infty}v_{n}\xi^{n},\quad U\left(
\xi\right)  =\sum_{n=1}^{\infty}u_{n}\xi^{n},\nonumber\\
\quad X\left(  \xi\right)   &  =\sum_{n=0}^{r-1}x_{n}\xi^{n}. \label{fourier}%
\end{align}
It then follows from Eq. (\ref{Wiener}) that we can get
\begin{equation}
D\left(  \xi\right)  X\left(  \xi\right)  =Y\left(  \xi\right)  +V\left(
\xi^{-1}\right)  +U\left(  \xi\right)  \xi^{r-1} \label{fourier form}%
\end{equation}
for $|\xi|=1$. Becuase $D\left(  \xi\right)  $ and $\ln D\left(  \xi\right)  $
is continuous and periodic on the unit circle, $D\left(  \xi\right)  $ has a
unique factorization, up to a multiplicative constant, in the form
\begin{equation}
D\left(  \xi\right)  =P^{-1}\left(  \xi\right)  Q^{-1}\left(  \xi^{-1}\right)
, \label{Dxi}%
\end{equation}
for $|\xi|=1$, such that $P\left(  \xi\right)  $ and $Q\left(  \xi\right)  $
are both analytic for $|\xi|<1$ and continuous and nonzero for $|\xi|\leq1$.
we may now use the factorization of $D\left(  \xi\right)  $ in Eq.
(\ref{fourier form}) to write
\begin{align}
&  P^{-1}\left(  \xi\right)  X\left(  \xi\right)  -\left[  Q\left(  \xi
^{-1}\right)  Y\left(  \xi\right)  \right]  _{+}\nonumber\\
&  \quad\quad\quad\quad\quad\quad-\left[  Q\left(  \xi^{-1}\right)  U\left(
\xi\right)  \xi^{r-1}\right]  _{+}\nonumber\\
&  =\left[  Q\left(  \xi^{-1}\right)  Y\left(  \xi\right)  \right]
_{-}+Q\left(  \xi^{-1}\right)  V\left(  \xi^{-1}\right) \nonumber\\
&  \quad\quad\quad\quad\quad\quad\quad\quad\quad+\left[  Q\left(  \xi
^{-1}\right)  U\left(  \xi\right)  \xi^{r-1}\right]  _{-}, \label{pq}%
\end{align}
where the subscript $+\left(  -\right)  $ means that we should expand the
quantity in the brackets into a Laurent series and keep only those terms where
$\xi$ is raised to a non-negative (negative) power. The left-hand side of Eq.
(\ref{pq}) defines a function analytic for $|\xi|<1$ and continuous on
$|\xi|=1$ and the right-hand side defines a function which is analytic for
$|\xi|>1$ and is continuous for $|\xi|=1$. Taken together they define a
function $E(\xi)$ analytic for all $\xi$ except possibly for $|\xi|=1$ and
continuous everywhere. But these properties are sufficient to prove that
$E(\xi)$ is an entire function which vanished at $|\xi|=\infty$ and thus, by
Liouville's theorem, must be zero everywhere \cite{McCoy Wu,McCoy}. Therefore
both the right-hand side and the left-hand side of Eq. (\ref{pq}) vanish
separately and thus we have
\begin{align}
X\left(  \xi\right)   &  =P\left(  \xi\right)  \left[  Q\left(  \xi
^{-1}\right)  Y\left(  \xi\right)  \right]  _{+}\nonumber\\
&  +P\left(  \xi\right)  \left[  Q\left(  \xi^{-1}\right)  U\left(
\xi\right)  \xi^{r-1}\right]  _{+}.
\end{align}
Furthermore, $U\left(  \xi\right)  $ can be neglected for large $r$%
\begin{equation}
X\left(  \xi\right)  \approx P\left(  \xi\right)  \left[  Q\left(  \xi
^{-1}\right)  Y\left(  \xi\right)  \right]  _{+}. \label{xxi}%
\end{equation}
Consider the term $\left[  Q\left(  \xi^{-1}\right)  Y\left(  \xi\right)
\right]  _{+}$, because $Q\left(  \xi\right)  $ is a $+$ function, so we can
expand it as a Laurent series and keep only those term where $\xi$ is raised
to a non-negative power,
\begin{align}
Q\left(  \xi\right)   &  ={\sum\limits_{n=0}^{\infty}}a_{n}\xi^{n}\nonumber\\
&  =\left(  a_{0}+a_{1}\xi^{1}+a_{2}\xi^{2}+\cdots+a_{r-1}\xi^{r-1}\right)
+O\left(  \xi^{r}\right)  ,
\end{align}
and then
\begin{equation}
Q\left(  \xi^{-1}\right)  =a_{0}+a_{1}\xi^{-1}+a_{2}\xi^{-2}+\cdots+a_{r-1}%
\xi^{1-r},
\end{equation}
where we have neglected the term $O\left(  \xi^{r}\right)  $\ for large $r$
for clarity, from Eq. (\ref{sumeq}) and Eq. (\ref{fourier}), we have
\begin{align}
Y\left(  \xi\right)   &  =\sum_{n=0}^{r-1}y_{n}\xi^{n}\nonumber\\
&  =\frac{x}{N}\left(  1+\operatorname{e}^{\operatorname*{i}k}\xi
^{1}+\operatorname{e}^{2\operatorname*{i}k}\xi^{2}+\cdots+\operatorname{e}%
^{\operatorname*{i}\left(  r-1\right)  k}\xi^{r-1}\right)  ,
\end{align}
thus
\begin{align}
&  \left[  Q\left(  \xi^{-1}\right)  Y\left(  \xi\right)  \right]
_{+}\nonumber\\
&  =\frac{x}{N}[\left(  a_{0}+a_{1}\operatorname{e}^{\operatorname*{i}k}%
+a_{2}\operatorname{e}^{2\operatorname*{i}k}+\cdots+a_{r-1}\operatorname{e}%
^{\operatorname*{i}\left(  r-1\right)  k}\right) \nonumber\\
&  \mathstrut\quad+\left(  a_{0}\operatorname{e}^{\operatorname*{i}k}%
+a_{1}\operatorname{e}^{2\operatorname*{i}k}+\cdots+a_{r-2}\operatorname{e}%
^{\operatorname*{i}\left(  r-1\right)  k}\right)  \xi^{1}\nonumber\\
&  \quad+\cdots+\left(  a_{0}\operatorname{e}^{\operatorname*{i}\left(
r-1\right)  k}\right)  \xi^{r-1}].
\end{align}
From Eq. (\ref{sumxn}), Eq. (\ref{fourier}) and Eq. (\ref{xxi}), we have%
\begin{align}
\sum_{n=0}^{r-1}\operatorname{e}^{-\operatorname*{i}kn}x_{n}^{(r-1)}  &
=X\left(  \operatorname{e}^{-\operatorname*{i}k}\right) \nonumber\\
&  =P\left(  \operatorname{e}^{-\operatorname*{i}k}\right)  \left[  Q\left(
\operatorname{e}^{\operatorname*{i}k}\right)  Y\left(  \operatorname{e}%
^{-\operatorname*{i}k}\right)  \right]  _{+},
\end{align}
and%
\begin{align}
&  \left[  Q\left(  \operatorname{e}^{\operatorname*{i}k}\right)  Y\left(
\operatorname{e}^{-\operatorname*{i}k}\right)  \right]  _{+}\nonumber\\
&  =\frac{x}{N}\left[  ra_{0}+ra_{1}\operatorname{e}^{\operatorname*{i}%
k}+ra_{2}\operatorname{e}^{2\operatorname*{i}k}+\cdots+ra_{r-1}%
\operatorname{e}^{\operatorname*{i}\left(  r-1\right)  k}\right] \nonumber\\
&  \mathstrut\quad-\frac{x}{N}\left[  a_{1}\operatorname{e}^{\operatorname*{i}%
k}+2a_{2}\operatorname{e}^{2\operatorname*{i}k}+\cdots+\left(  r-1\right)
a_{r-1}\operatorname{e}^{\operatorname*{i}\left(  r-1\right)  k}\right]
\nonumber\\
&  =\frac{x}{N}\left[  rQ\left(  \operatorname{e}^{\operatorname*{i}k}\right)
-\operatorname{e}^{\operatorname*{i}k}\frac{dQ\left(  \xi\right)  }{d\xi
}|_{\xi=\operatorname{e}^{\operatorname*{i}k}}\right]  . \label{qy}%
\end{align}
So when $r\gg1$, we can ignore the second term in Eq. (\ref{qy}). Together
with Eq. (\ref{Dxi}), we get
\begin{equation}
X\left(  \operatorname{e}^{-\operatorname*{i}k}\right)  =\frac{xr}{N}P\left(
\operatorname{e}^{-\operatorname*{i}k}\right)  Q\left(  \operatorname{e}%
^{\operatorname*{i}k}\right)  =\frac{xr}{ND\left(  \operatorname{e}%
^{-\operatorname*{i}k}\right)  }.
\end{equation}
At last, by Szeg\"{o}'s theorem, we get
\begin{equation}
\Delta_{r}=\mu^{r}\exp(\sum_{n=1}^{\infty}nd_{-n}d_{n}),
\end{equation}
where%
\begin{align}
\mu &  =\exp\left[  \int_{-\pi}^{\pi}\frac{dq}{2\pi}\,\ln D(\operatorname{e}%
^{\operatorname*{i}q})\right]  ,\nonumber\\
d_{n}  &  =\int_{-\pi}^{\pi}\frac{dq}{2\pi}\,e^{-\operatorname*{i}qn}\ln
D(\operatorname{e}^{\operatorname*{i}q}).
\end{align}
From Eq. (\ref{sumxn}), we have%
\begin{equation}
\Theta(r,N,x,\operatorname{e}^{\operatorname*{i}k})=\Delta_{r}\left(
1+\frac{xr}{ND\left(  \operatorname{e}^{-\operatorname*{i}k}\right)  }\right)
.
\end{equation}
\textbf{Q.E.D.}

\section{Evaluation of the correlation functions}

Now we can evaluate $\Gamma(r,N,\alpha_{k},\operatorname{e}^{\operatorname*{i}%
k})$ in Eq. (\ref{Gamma}) by applying the \emph{Theorem} directly to get%
\begin{align}
\Gamma(r,N,\alpha_{k},\operatorname{e}^{\operatorname*{i}k})  &
=\Gamma(r,N,\alpha_{-k},\operatorname{e}^{-\operatorname*{i}k})\nonumber\\
&  =\left(  -1\right)  ^{r}\left(  1-\frac{h^{2}}{J^{2}}\right)  ^{1/4}\left(
1-\frac{2r}{N}\right)  ,
\end{align}
for large $r$ and $N$. So the asymptotic behavior of the correlation functions
is given by ($k\in q^{(o)}\cup q^{(e)}$)
\begin{align}
  C_{r,N}^{xx}\left(  |E_{k}^{(o/e)}\rangle\right) =\left(  -1\right)  ^{r}\left(  1-\frac{h^{2}}{J^{2}}\right)  ^{1/4}\left(1-\frac{2r}{N}\right),
\end{align}
which is independent of $k$. Thus, the correlation functions of the $2N$
low-lying states possess the same asymptotic behavior. If assuming the
cannonical ensemble for the low-lying energy levels, one would agree that the
$2N$ levels dominate the system's properties at low temperatures ($T$
$\ll4h/k_{B}$, $k_{B}$ is the Boltzmann constant) and arrive at a conclusion
that the thermal correlation function is inert to temperature meanwhile.

This work was supported by the NSFC under Grants no. 11074177, SRF for ROCS
SEM (20111139-10-2).

\end{document}